%% file: main.tex
\documentclass[aps, prb,  reprint,  superscriptaddress,  longbibliography,  showkeys,  citeautoscript,]{revtex4-2}
\input{settings/settings}

\makeatletter
\newcommand\emailx[1]{%
\move@AF%
\def\@affil{{\normalfont\,#1\strut}{}}%
}%

\begin{document}

\title{Path encoding entanglement distribution in multicore fiber using photonic integrated circuits}
\title{Chip-to-chip entanglement distribution over 80-km multicore fiber link}
%\title{Two-node integrated photonic quantum network for long-distance entanglement distribution}

\input{settings/authors}


\begin{abstract}
\input{text/abstract}
\end{abstract}
\maketitle

\section{Introduction}
\input{text/intro}

\section{Results}
\input{text/results_exp_setup}
\input{text/results_qst}
\input{text/results_skr}

\section{Discussion}
\input{text/discussion}

\section{Conclusion}
\input{text/conclusion}

\section{Methods}
\input{text/methods}

\input{text/ackowledgments}

\input{text/contributions}

\section*{Competing Interests}
The authors declare they have no competing interests.

\section*{Data availability statement}
Data supporting the findings of this study are available within the article.
%and Supplementary Material
Raw data and supplementary information are available from the corresponding authors upon request.

\section*{References}
\bibliography{bibliography}
%TC:endignore
\appendix

\end{document}

%% file: settings/settings.tex
\usepackage{silence}
\usepackage[intlimits]{amsmath}
\usepackage{amssymb}
\usepackage{amsthm}
\usepackage{booktabs}
\usepackage{braket}
\usepackage{color}
\usepackage{graphicx}
\usepackage{float}

\usepackage[colorlinks=true,
            linkcolor=black,
            citecolor=black,
            filecolor=black,
            urlcolor=black
            ]{hyperref}
\usepackage[utf8]{inputenc}
\usepackage[version=4]{mhchem}
\usepackage{multirow}
\usepackage{soul}
\usepackage{tabularx}
\usepackage{units}
\usepackage[dvipsnames]{xcolor}

\usepackage{dcolumn}
\usepackage{booktabs}
\usepackage{miller}
\usepackage{sidecap} % For Figure 1 side caption

\makeatletter
\def\@hangfrom@section#1#2#3{\@hangfrom{#1#2}#3}%\MakeTextUppercase{#3}}%
\def\@hangfroms@section#1#2{#1#2}%\MakeTextUppercase{#2}}%
\makeatother

\setcitestyle{super}

%% file: settings/authors.tex
\newcommand{\adrdtu}{Department of Electrical and Photonics Engineering, Technical University of Denmark, Ørsteds Pl., Kgs. Lyngby, 2800, Denmark}
\newcommand{\adrflo}{Department of Physics and Astronomy, University of Florence, Sesto Fiorentino, 50019, Italy}
\newcommand{\adrsiphotonics}{SiPhotonIC ApS, Virum, 2830, Denmark}

\author{Damien Roux}
\altaffiliation{Contributed equally to this work}
%\email{damro@dtu.dk}
\affiliation{\adrdtu}

\author{Giulia Guarda}
\altaffiliation{Contributed equally to this work}
%\email{giulia.guarda@unifi.it}
\affiliation{\adrflo}

\author{Mujtaba Zahidy}
\affiliation{\adrdtu}

\author{Yunhong Ding}
\affiliation{\adrdtu}
\affiliation{\adrsiphotonics}

\author{Siyan Zhou}
\affiliation{\adrsiphotonics}

\author{Domenico Ribezzo}
\affiliation{\adrflo}

\author{Battulga Munkhbat}
\affiliation{\adrdtu}

\author{Francesco Da Ros}
\affiliation{\adrdtu}

\author{Davide Bacco}
\affiliation{\adrflo}

\author{Caterina Vigliar}
\email{catvi@dtu.dk}
\affiliation{\adrdtu}

%% file: text/abstract.tex
Long-range quantum entanglement is essential for building large-scale quantum networks and unconditionally secure cryptographic systems based on quantum key distribution (QKD). 
While photonic integrated circuits offer a highly scalable platform, the fragility of phase coherence between spatial modes has prevented the distribution of path-encoded entanglement over long distances.
Here, we report chip-to-chip distribution of path-encoded entangled states over 80 km between fully integrated silicon photonic transmitter and receiver chips.
Telecom-band entangled photon pairs are generated via spontaneous four-wave mixing in on-chip spiral waveguides and distributed between chips over a dual-core, actively stabilized fiber link.
Upon distribution, we measure a Bell state fidelity of $85.7 \pm 0.2 \%$.
Implementing the BBM92 protocol with the same source, we obtain a secure key rate of 2.03 bit/s in the infinite-key regime.
These results establish silicon photonic chips as a viable platform for long-distance path-encoded entanglement-based quantum key distribution, paving the way toward scalable, device-independent quantum networks.

%% file: text/intro.tex
The ability to distribute entanglement across long distances is fundamental for building large-scale quantum networks \cite{wehner2018quantum}, enabling applications such as quantum communication, quantum sensing, and distributed quantum computing \cite{azuma2023quantum, krenn2017entanglement}. Among these, quantum key distribution (QKD) stands out as the most mature and technologically advanced quantum communication protocol to date \cite{pirandola2020advances, scarani2009security, wehner2018quantum}, offering two distant parties the ability to establish a shared secure cryptographic key with unconditional security \cite{xu2020secure}. 
Unlike prepare-and-measure schemes based on weak coherent pulses, entanglement-based QKD protocols remove the need to trust the source,
mitigate vulnerabilities to photon-number-splitting attacks, and are naturally compatible with quantum repeaters and entanglement swapping.
These capabilities are essential for device-independent (DI) security and for extending quantum communication beyond point-to-point transmission distances, toward a global quantum internet \cite{pirandola2020advances, azuma2023quantum}. However, entanglement-based demonstrations have largely employed bulk optical setups \cite{yin2020entanglement, wengerowsky2019submarine, neumann2022continuous}, limiting their practicality and prospects for large-scale deployment.

Photonic integrated circuits (PICs) offer a scalable and flexible approach for implementing such protocols \cite{wang2020integrated}. 
Their compatibility with existing telecom infrastructure, inherent phase stability, and support for standard integrated components make them well-suited for manipulating quantum states \cite{bao2023very, krenn2017entanglement, wang2018multidimensional}. Path encoding, where a quantum bit is encoded in the superposition of two spatial modes, is the most natural encoding for on-chip implementations \cite{metcalf2014quantum, politi2008silica, shadbolt2012generating, silverstone2014chip, silverstone2015qubit, tanzilli2012genesis}, offering direct manipulation via integrated Mach-Zehnder interferometers (MZIs) \cite{bogaerts2020programmable}. 

Current experimental efforts in PIC-based QKD address the practical challenges that still hinder the large-scale deployment of QKD systems i.e., increasing transmission rates, extending achievable distances, and advancing toward device-independent security.
Despite the advantages of PICs, practical systems have primarily relied on weak coherent pulses using various encoding schemes, including polarization \cite{ma2016silicon}, time-bin \cite{semenenko2020chip, sibson2017chip, sibson2017integrated}, and path \cite{ding2017high}, but entanglement-based implementations remain comparatively limited. 
While recent demonstrations have exploited frequency-bin \cite{khodadad2025frequency, tagliavacche2025frequency} and time-bin \cite{yu2025quantum} entanglement for on-chip QKD, 
achieving comparable secure key rates with path-encoded states remains challenging due to the fragility of phase coherence between spatial modes, which is highly sensitive to environmental perturbations, which become harder to compensate at longer distances \cite{bacco2021mcf}.

To date, experimental progress in chip-to-chip entangled state distribution has been limited to three path-encoded demonstrations.
Two of these converted path-encoding to polarization using bidimensional grating couplers for transmission \cite{wang2016chip, llewellyn2020chip}, which limits the state to two dimensions and caps the scalability advantage of path encoding. The third short multicore fiber but interleaved the stabilization with quantum measurements, resulting in slow correction cycles incompatible with the phase drifts experienced over long fiber links \cite{thomas2025high}.
Notably, demonstrating that path-encoded entanglement can be distributed over meaningful distances without conversion and exploited for QKD remains an open milestone for the community.

In this work, we address this gap by demonstrating a proof-of-concept entanglement-based QKD system using fully integrated silicon photonic transmitter and receiver chips connected by fiber links up to 80 km (Figure~\ref{Fig:concept}). Telecom-band entangled photon pairs are generated via spontaneous four-wave mixing (SFWM) in centimeter-scale spiral waveguides on the transmitter chip (Alice) to produce a spatially encoded entangled state \cite{adcock2019programmable, llewellyn2020chip, paesani2019generation, vigliar2021error, zhang2019generation}. The photon pair is split between the two nodes: the idler photon remains on Alice's chip, while the signal photon is transmitted to the receiver chip (Bob) via a multicore fiber (MCF). 
Because the two cores of the MCF experience highly correlated environmental noise \cite{bacco2021mcf}, the relative phase fluctuations between the signal paths remain slow. 
We dynamically compensate these drifts using a phase-locked loop (PLL) that exploits residual pump light co-propagating with the signal photons as a phase reference, enabling continuous phase tracking  without the need for additional lasers. 
The slow differential drift in the MCF falls within the correction bandwidth of the PLL, making long-distance operation possible \cite{da2019stable}.

To perform state characterization and QKD protocols, local measurement bases are selected by applying unitary transformations to the dual-rail encoded qubits on Alice's and Bob's PICs, respectively. Photon correlations are recorded using single-photon detectors.
We investigate the entanglement quality across channel links of increasing length, quantifying the degradation induced by propagation loss and phase decoherence. Building on these measurements, we implement QKD and demonstrate secure key exchange over such links.
In an initial characterization, the two chips are connected via approximately 4 m of single-mode fiber, achieving a secure key rate (SKR) of \(\mathrm{802.57}~\mathrm{bit/s}\).
In a second, long-distance demonstration over an \(80~\mathrm{km}\) two-core fiber spool, we achieve a Bell pair fidelity  of $\mathcal{F} = 85.7 \pm 0.2 \% $ 
and a SKR of \(\mathrm{2.03}~\mathrm{bit/s}\).
These results demonstrate that entanglement fidelity can be preserved over inter-city distances while supporting path-encoded QKD across practically relevant links, marking a step toward the interconnection of integrated devices in future quantum networks.

%% file: text/results_exp_setup.tex
\subsection{Experimental setup}

\begin{figure*}[t!]
    \includegraphics[width=1\linewidth]{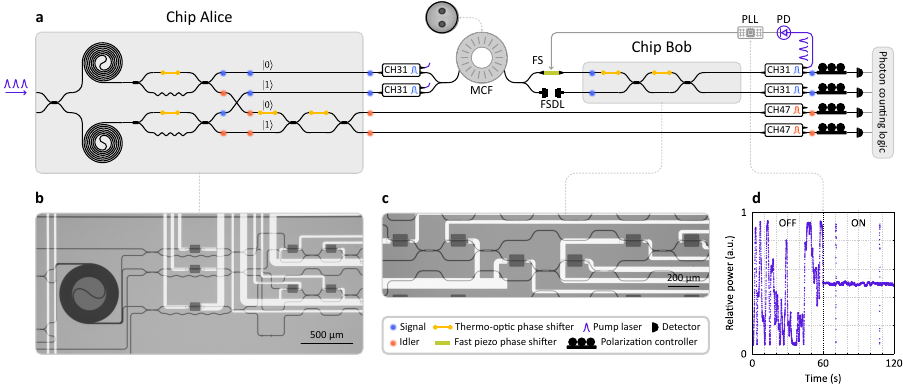}
    \centering
    \caption{
    \textbf{Experimental setup and integrated photonic chips.}
    \textbf{a}
    Alice’s silicon photonic transmitter (left) generates telecom-band entangled photon pairs in two spiral waveguides. Asymmetric MZIs spatially separate the signal and idler photons of different colors, encoding them in the path degree of freedom. Alice performs projective measurements on the idler photons locally, while the signal photons are transmitted to Bob's receiver chip (right) via a multicore fiber (MCF). Both parties perform complementary unitaries and route the photons to superconducting nanowire single-photon detectors (SNSPDs). 
    Dense wavelength division multiplexing (DWDM) filters separate the signal (channel 31) and the idler (channel 47) from the pump. 
    Before the MCF, an additional set of DWDMs on the signal paths suppresses the pump. After the receiver chip, another set filters out noise photons generated via pump-MCF interactions and isolates the residual pump power required for phase stabilization.
    A photodiode (PD) monitors this power, feeding a phase-locked loop (PLL) system that controls an external fiber stretcher (FS) to stabilize the relative phase between the signal paths during key distribution. A free-space delay line (FSDL) ensures path-length matching with subwavelength precision, maximizing two-photon interference visibility.
    \textbf{b, c} Optical microscope images of the transmitter and receiver chips, respectively. Scale bars as indicated.  
    \textbf{d}
    Optical power recorded by the photodiode over time, with and without active phase stabilization.
    Without stabilization, environmental perturbations induce random relative phase fluctuations in the inter-chip link, even in multicore fibers, on a timescale of seconds.
    With active locking, the feedback loop compensates for
    % drifts up to the few-Hz range, stabilizing
    the slow, sub-Hz fluctuations in the multicore link. Brief power jumps are phase unlocks triggered by sudden perturbations, which are rapidly corrected.
    }
    \label{Fig:path_setup_complete}
\end{figure*}

A schematic of the experimental setup is shown in Figure \ref{Fig:path_setup_complete}a. A pulsed pump laser is coupled into the integrated silicon transmitter chip via grating couplers. The pump is split on-chip to coherently drive two identical 1.5 cm-long spiral waveguides, each generating non-degenerate signal and idler photon pairs through SFWM. Operating in the low-gain regime suppresses simultaneous multi-pair emission across the spirals. On-chip filters then spectrally separate the signal and idler photons from each other. Since photon pairs generated in the first and second spirals (Figure \ref{Fig:path_setup_complete}a) encode the logical states $\ket{0}$ and $\ket{1}$ respectively, the device prepares a coherent superposition of these two spatial modes, yielding the maximally entangled path-encoded Bell state $\ket{\Phi^+} = \frac{1}{\sqrt{2}} (\ket{00}_{s,i} + \ket{11}_{s,i})$ \cite{silverstone2014chip}.

To characterize the distributed state, single-qubit projective measurements are performed by applying local unitary transformations to the idler and signal modes and single-photon detection. These are implemented using reconfigurable MZI meshes on the transmitter and receiver chips (Figure \ref{Fig:path_setup_complete}b,c), which act as tunable beam splitters to mix the spatial modes and apply phase shifts, enabling arbitrary single-qubit rotations.
The resulting four output spatial modes, two for the signal and two for the idler, are then coupled off-chip and routed to a detection system. 

We characterize the distribution in three configurations of increasing complexity: an on-chip baseline setup where both signal and idler photons are locally measured on the transmitter device, a chip-to-chip configuration where transmitter and receiver chips are connected by short single-mode fibers, and finally a setup with transmission over an 80 km dual-core fiber.

To enable chip-to-chip distribution, a free-space delay line (FSDL) is placed in one of the signal paths to compensate for a small optical path-length mismatch and ensure the temporal overlap necessary for the interference at the receiver (Figure \ref{Fig:path_setup_complete}a), and active phase stabilization continuously corrects the phase drifts introduced by the fiber (Figure \ref{Fig:path_setup_complete}d). The phase reference is provided by residual pump light that leaks through the DWDM filter and co-propagates with the signal photons through the fiber, acquiring the same phase noise. After the receiver chip, the pump is separated from the signal photons and detected by a high-responsivity photodiode (PD). This signal feeds a phase-locked loop (PLL) that drives a fiber stretcher (FS) placed on one of the two signal modes, effectively stabilizing the differential phase in both short- and long-range configurations.

\begin{figure}[t]
    \centering
    \includegraphics[width = 0.48\textwidth]{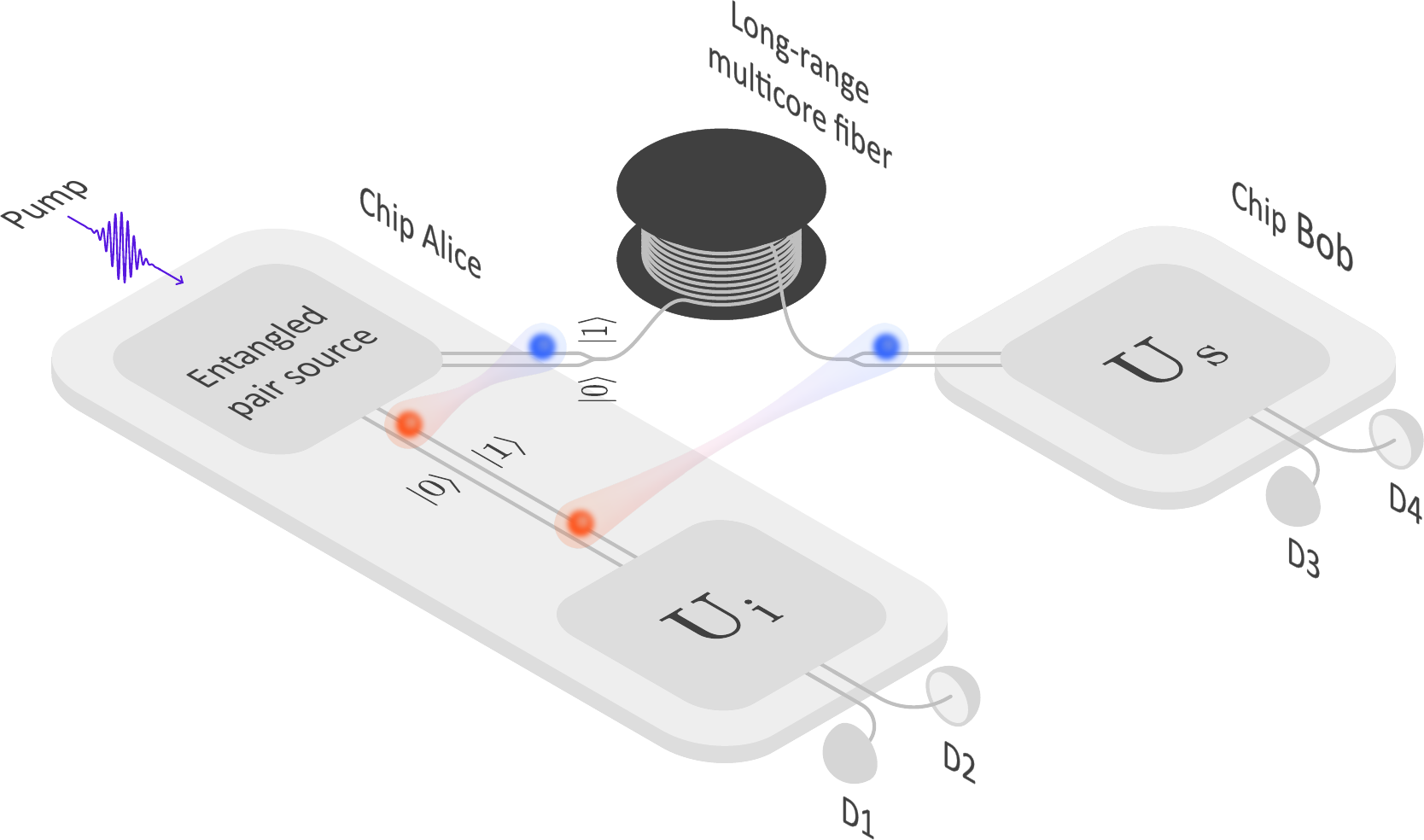}
    \caption{\textbf{Conceptual scheme of the chip-to-chip quantum state distribution experiment.} An external pump laser excites two integrated sources on the transmitter chip (Alice). The generated idler photons are directed to Alice's local measurement station, while the signal photons are distributed over a multicore fiber to the receiver chip (Bob).
    For each qubit, measurement bases are chosen by applying unitary transformations ($U_i$ and $U_s$), followed by detection for two-photon coincidence event counting.}
    \label{Fig:concept}
\end{figure}

%% file: text/results_qst.tex
\subsection{Quantum state tomography}

\begin{figure*}[t!]
    \centering
    \includegraphics[width=1\textwidth, trim={0pt 0pt 0pt 0pt}, clip]{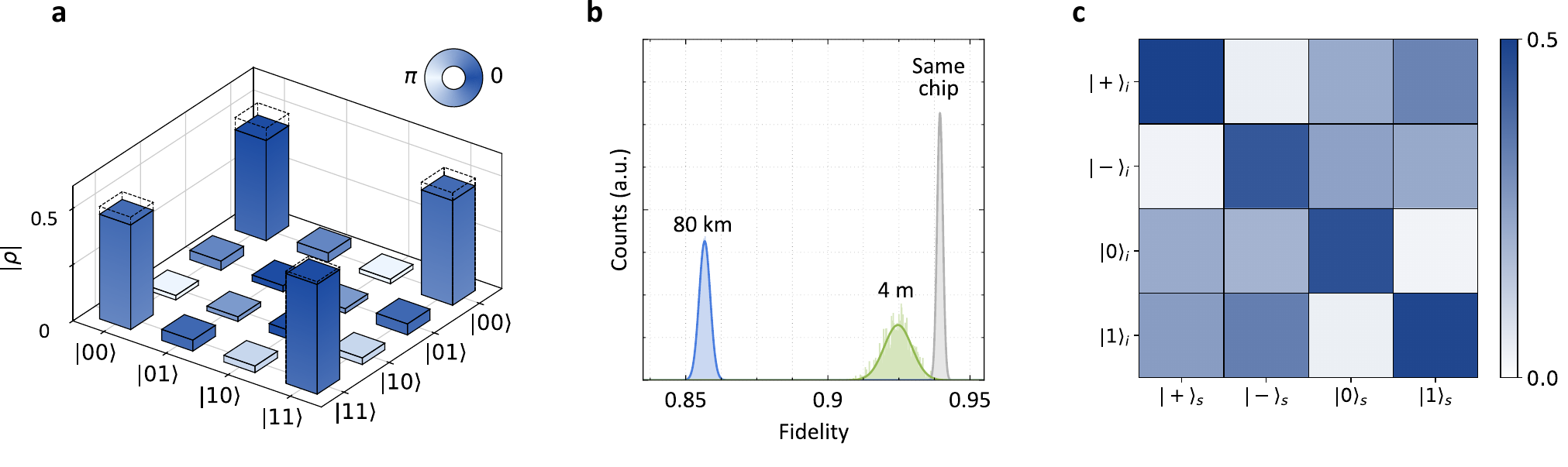}
    \centering
    \caption{
    \textbf{Quantum state tomography}. 
    \textbf{a} Reconstructed density matrix
    %$\rho_{\textup{exp}}$ 
    of the photon-pair state after sending the signal photons to Bob's chip through the 80 km dual-core fiber. The height of the bars denotes the amplitude of the matrix entries 
    %$\rho_{\text{exp}_{ij}}$ 
    and their color encodes the phase. The theoretical density matrix of the Bell state $\ket{\Phi^+}$ is shown as a dotted line. 
    \textbf{b} Fidelity histograms obtained from Monte Carlo simulations for Alice and Bob connected by 4 m single-mode fibers and an 80 km multicore fiber. The single-chip case, where Alice and Bob are located on the same chip, is displayed for reference.
    For each run, new coincidence datasets are sampled from the estimated noise distribution and the fidelity is recomputed from the new density matrix. The histogram shows the resulting distribution over 20 000 runs. Noise is estimated from the coincidence count statistics (same chip, 4 m), or modeled as Poissonian (80 km).
    \textbf{c} Joint measurement outcome probabilities in the mutually unbiased bases $Z$ and $X$ for the signal and idler photons, measured over the 80 km dual-core fiber link.}
    \label{fig:tomo_cph}
\end{figure*}

We perform full tomography to evaluate the entanglement quality by projecting the signal and idler photons onto a complete set of bases defined by the Pauli operators: the computational basis ($Z$), and the superposition bases ($X$, and $Y$). The density matrices are reconstructed via maximum likelihood estimation \cite{James2001}, and capture the end-to-end performance of the system, including the quality of preparation, distribution, and measurement. 
The matrix obtained with the 80 km link is shown in Figure~\ref{fig:tomo_cph}\textbf{a}, in good agreement with the target Bell state $\ket{\Phi^+} \bra{\Phi^+}$. To quantify this agreement, we compute the state fidelity $\mathcal{F}$, with uncertainties estimated via Monte Carlo simulation (see Methods). The resulting fidelities are presented in Figure~\ref{fig:tomo_cph}\textbf{b}, with $\mathcal{F} = 94.0 \pm 0.1\% $ for the single-chip case,  $\mathcal{F} = 92.5 \pm 0.5\% $ for the short link connecting the two PICs, and $\mathcal{F} = 85.7 \pm 0.2 \% $ using the long multicore fiber. In the latter case, the state yields a maximum CHSH (Clauser–Horne–Shimony–Holt) value of $S = 2.648 \pm 0.004$, in clear violation of the classical bound $S \leq 2$, by 166 standard deviations. 

From a subset of the full tomography data, we study the joint measurement probabilities in the $Z$ and $X$ bases. The results are presented in Figure~\ref{fig:tomo_cph}\textbf{c} as a $4 \times 4$ matrix composed of four normalized $2 \times 2$ subspaces corresponding to the joint observables $XX$, $XZ$, $ZX$, and $ZZ$ \cite{wang2018multidimensional, tagliavacche2025frequency}. 
In an ideal scenario,  outcomes should show perfect correlation when both parties select the same measurement basis ($XX$ and $ZZ$), and yield uniformly random, uncorrelated results when different bases are chosen ($XZ$ and $ZX$), as expected for two mutually unbiased bases. 
Our experimental results over the 80 km link closely follow this expected pattern, with an overlap with the theoretical matrix of 97.9\%.

%% file: text/results_skr.tex
\subsection{QKD protocol and key rate analysis}

We analyze the secure key exchange using the BBM92 protocol, an entanglement-based variant of BB84 \cite{bennett1992quantum}. This approach ensures security even with an untrusted source positioned between Alice and Bob \cite{pirandola2020advances}. Furthermore, the use of entanglement relaxes the constraints in the underlying security proofs \cite{scarani2009security, pirandola2020advances}.

Following the procedure described in the Methods section, we evaluate the quantum bit error rate (QBER) and the corresponding SKR for a distribution across the short link and the 80 km two-core fiber.
Table \ref{Table:qber_summary} summarizes the measured QBER across the measurement bases for both links.
We measure the QBER in all three measurement bases and identify the $Z$- and $X$-basis combination as the optimal configuration for QKD, which we then use to perform the secure key rate analysis.
As anticipated, the $X$ and $Y$ bases consistently show higher error rates than the $Z$ basis ($QBER_{X}, \, QBER_{Y} > QBER_Z$), 
as they require high-contrast interference between paths and are therefore more sensitive to the phase noise, which becomes increasingly difficult to suppress with transmission distance.

\begin{table}[h!]
    \centering
    \small
    \setlength{\tabcolsep}{4pt}
    \renewcommand{\arraystretch}{1.2}
    \begin{tabular}{lccc}
        \toprule
        Length & $\mathrm{QBER}_{Z}$ (\%) & $\mathrm{QBER}_{X}$ (\%) & $\mathrm{QBER}_{Y}$ (\%) \\
        \midrule
        4 m   & 2.58 & 5.60 & 7.27 \\
        80 km & 6.81 & 7.26 & 8.63 \\
        \bottomrule
    \end{tabular}
    \caption{QBER values for the $Z$, $X$, and $Y$ bases for the infinite key analysis.}
    \label{Table:qber_summary}
\end{table}

\begin{figure}[t]
    \centering
    \includegraphics[width=0.3\textwidth, trim=0 0 0 0, clip]{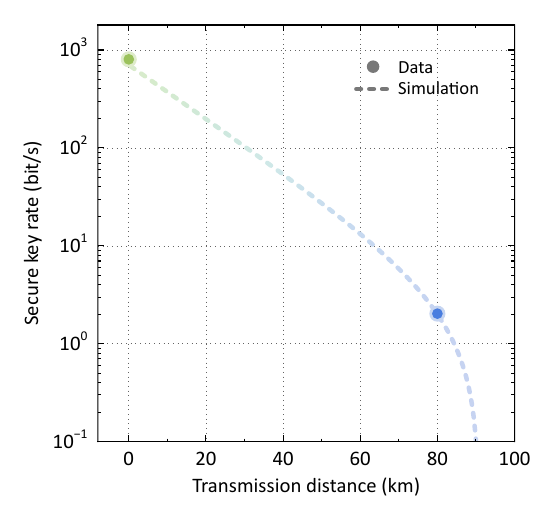}
    \caption{\textbf{Secure key rate (SKR) as a function of fiber length.} The measured values at 4 m (green) and 80 km (blue) closely follow the simulated theoretical trend (dashed curve). As established, the SKR was evaluated using the optimal $Z$ and $X$ measurement bases.}
    \label{fig:skr_results}
\end{figure}

Figure \ref{fig:skr_results} shows the calculated SKR for the two links, yielding 803 bit/s for the short link and 2.03 bit/s for the two-core fiber. Both values are in strong agreement with the theoretical simulation (dashed curve), which incorporates source rates, losses and detector parameters (see Supplementary Information).
We also investigate the impact of finite-size effects on the SKR evaluation. In particular, Table \ref{Table:cph_skr_finite} reports the SKR for different block sizes $n$, showing that the extractable key rate remains robust down to $n=10^5$, with only a modest reduction compared to the asymptotic limit. 
The maximum transmission distance is ultimately constrained by the coupling efficiency of the chips ($\sim 7$ dB per grating coupler), which accounts for the majority of the losses.

\begin{table}[h!]
    \centering
    \setlength{\tabcolsep}{4pt}
    \renewcommand{\arraystretch}{1.2}
    \begin{tabular}{lrrrr}
        \toprule
        Length & $n = 10^8$ & $n = 10^7$ & $n = 10^6$ & $n = 10^5$ \\
        \midrule
        4 m   & 802 & 800 & 794 & 775 \\
        80 km & 2.03  & 2.02   & 1.98   & 1.87   \\
        \bottomrule
    \end{tabular}
    \caption{SKR values (bit/s) for different block sizes $n$. 
    }
    \label{Table:cph_skr_finite}
\end{table}

%% file: text/discussion.tex
\begin{table*}[t]
\centering
\small
\setlength{\tabcolsep}{8pt}
\renewcommand{\arraystretch}{1.2} 
\begin{tabular}{llcccc}
\toprule
Ref & Source & Encoding & QBER (\%) & SKR (bps) & Distance (km) \\
\midrule
\citenum{liu2023photonic} & Si waveguide (spiral) & Time-energy & 7 & 4 & 12.3 \\
\citenum{wen2022realizing} & Si\textsubscript{3}N\textsubscript{4} (micro-ring resonator) & Time-energy & 3.08 & 205 & 0 \\
\citenum{yu2025quantum} & SiO\textsubscript{2} waveguide (spiral) & Time-bin & -- & 200 & 25 \\
\citenum{tagliavacche2025frequency} & Si (two micro-ring resonators) & Frequency & 10.4 & 4.5 & 26 \\
\midrule
This work & Si waveguide (spiral) & Path & 4.1 & 2.03 & 80 \\
\bottomrule
\end{tabular}
\caption{Comparison of secret key rate (SKR) and average QBER (between Z and X bases) reported in the literature for different integrated entangled photon-pair sources and encoding schemes.}
\label{Table:comparison_path}
\end{table*}

We report long-distance distribution of path-encoded entangled states between integrated silicon photonic chips.
Table \ref{Table:comparison_path} benchmarks our results against other integrated entanglement-based QKD systems, highlighting that our work achieves the longest transmission distance reported to date, with a key rate comparable to demonstrations over much shorter distances.

While this demonstration validates the feasibility of integrated path-encoded QKD, several hardware improvements could significantly enhance the performance of the system.
The main limitation arises from the high fiber-chip coupling losses ($\sim$7 dB per facet), which could be significantly reduced by engineering highly efficient couplers \cite{ding2014fully}.
A second bottleneck is the low speed of the integrated therm-optic phase shifters ($\sim 5 \ \mathrm{kHz}$),  which constrains the basis switching rate and ultimately the SKR.
Replacing these with electro-optic modulators would enable significantly higher modulation speeds, achievable either natively in silicon \cite{Thomson2011}, or through heterogeneous integration with lithium niobate (LiNbO\textsubscript{3}) \cite{wang2024ultrabroadband, xu2020high} or indium phosphide (InP) \cite{sibson2017chip, sibson2017integrated}.
Balanced beam splitters could also be used for passive randomized basis selection, eliminating the need for active basis switching at the cost of additional detectors and calibration complexity.
Additionally, integrating the phase-locked loop directly onto the photonic chip, to control an integrated phase shifter, rather than an external bulk modulator, would further reduce the system footprint and improve scalability.
The required photodetectors, phase shifters and feedback electronics can be readily co-integrated on the same substrate using state-of-the-art monolithic electronic-photonic platforms \cite{sun2015microprocessor, atabaki2018integrating}, where closed-loop stabilization and automated tuning have already been demonstrated \cite{Grillanda2014}, or alternatively realized through co-packaged systems \cite{margalit2021perspective}.

While polarization encoding is better suited for free-space and satellite communications \cite{ren2017ground, yin2017satellite}, and time-bin encoding offers robustness in fiber-based networks \cite{bacco2019field, guarda2023bb84, ribezzo2022deploying, fitzke2022scalable}, path encoding at telecom wavelengths is compatible with existing telecommunication infrastructure, including multicore fiber networks actively being adopted for classical high-capacity transmission \cite{jorgensen2022petabit, Mateo2024, bacco2021mcf}, facilitating the coexistence of classical and quantum signals \cite{da2021path}.
Path encoding is also a natural choice for realizing high-dimensional quantum states (qudits) in integrated photonic platforms \cite{ding2017high, chi2022programmable, da2021path}, a feature that enhances both noise tolerance and information capacity per photon \cite{cozzolino2019high}. Experimental demonstrations have already reached dimensions up to 16 \cite{wang2018multidimensional}, and suggest that this approach is readily compatible with the architecture presented here.

Finally, the choice of photon source and material platform sets the performance ceiling of the system.
Probabilistic photon pair sources fundamentally limit the achievable state fidelity due to low, although non-negligible multi-pair emission.
This could partially be addressed by adopting micro-ring resonator sources, which improve spectral purity and remove the need for lossy narrowband filtering required with standard waveguide sources \cite{llewellyn2020chip}, or intermodal SFWM, which simultaneously achieves high purity, heralding efficiency and indistinguishability \cite{paesani2020near}.
Future implementations could mitigate this through deterministic single-photon sources \cite{Zahidy2024} or improved heralding schemes \cite{Xiong2016, Davis2022}. 
Furthermore, while silicon is a well-established platform for SFWM \cite{adcock2019programmable, liu2023photonic, llewellyn2020chip, metcalf2014quantum, paesani2019generation, politi2008silica, shadbolt2012generating, silverstone2014chip, silverstone2015qubit, tanzilli2012genesis, vigliar2021error, zhang2019generation}, it suffers from two-photon absorption at high pump powers. Silicon nitride (Si\textsubscript{3}N\textsubscript{4}) provides lower on-chip and coupling losses \cite{ramelow2015silicon, xiang2022silicon}.
Alternatively, spontaneous parametric down-conversion (SPDC) sources in periodically poled lithium niobate (PPLN) waveguides \cite{fitzke2022scalable} or III–V semiconductors such as aluminum gallium arsenide (AlGaAs) \cite{appas2021flexible, autebert2016multi, steiner2023continuous} and gallium arsenide (GaAs) offer higher nonlinear coefficients. However, they require strict phase-matching conditions, making them less practical for field-trial applications. 
Ultimately, silicon remains a highly practical platform for near-term deployment, and the improvements outlined above provide a clear path toward the key rates required for real-world quantum networks.

%% file: text/conclusion.tex
In summary, we have demonstrated the chip-to-chip distribution of path-encoded entangled pairs over an 80-km fiber link between integrated silicon photonic chips, achieving a state fidelity of $85.7 \pm 0.2\%$ and a secure key rate of 2.03 bit/s. Active phase stabilization, exploiting residual pump light co-propagating with the single photons, ensures robust operation over the fiber link without any additional optical resource. By showing that high-fidelity path-entanglement can be preserved over long fiber distances, this architecture paves the way toward device-independent quantum networks, where security is guaranteed by Bell inequality violations alone.

%% file: text/methods.tex
\subsection{Measurements and QST analysis}

A mode-locked laser with 10 GHz repetition rate is fed to an intensity modulator, allowing the repetition rate to be reduced down to 50 MHz via pulse picking, for photon pair generation.
Light is coupled in and out of the chips via fiber arrays coupled to the on-chip grating couplers. After initial calibration of the on-chip filters and MZI meshes, the pump power is adjusted to suppress multi-pair emission to $\simeq 3\%$ for both spirals. The two spirals are balanced using on-chip tunable MZIs acting as attenuators to equalize their respective single-photon emission rates. 

Precise path-length alignment between the two chips is required to ensure temporal overlap between the distributed photons. This is achieved using a free-space delay line, adjusted to maximize the interference visibility of classical pump pulses transmitted through the fiber link.
In particular, when using the 80 km two-core fiber, the pump must be suppressed before the fiber to avoid nonlinear effects and the generation of additional spurious photons in the multicore fiber, while keeping enough power to operate the phase-locked loop. To satisfy both requirements, we use an avalanche photodiode, and its electrical signal is amplified before the PLL. Alternatively, a CW laser may be introduced to reach the minimum power required by the photodetector. The PLL stabilizes the recorded power to a user-defined setpoint, with its PID parameters optimized beforehand for maximum stability.

The projective measurements in the 
$Z$ ($\{\ket{0},\ket{1}\}$),
$X$ ($\ket{\pm}=1/\sqrt{2}(\ket{0}\pm \ket{1})$) , 
and $Y$ ($\ket{\pm i}=1/\sqrt{2}(\ket{0}\pm i \ket{1})$) 
bases are performed by tuning the MZI phases. The measurement procedure is automated, and active feedback guarantees phase stability throughout data collection.
State tomography is performed by sequentially applying all 9 combinations of signal and idler projection bases. The density matrix is reconstructed using $\rho = \frac{1}{4} \sum_{i,j=0}^{3} c_{ij} (\sigma_i \otimes \sigma_j)$, where $\sigma_i$ and $\sigma_j$ are the Pauli matrices, and $c_{ij}$ are real coefficients representing the expectation values of the joint observables. 
Coincidence counts are obtained via post-processing of acquired timestamps over an integration window.
Uncertainties are estimated via Monte Carlo simulation using Poissonian statistics on the coincidence counts.
For each Monte Carlo simulation, new coincidence datasets are generated by sampling from the estimated noise distribution, and the fidelity
\mbox{$\mathcal{F}(\rho_\textup{exp}, \rho_\textup{th}) = \left(\text{Tr}\sqrt{\sqrt{\rho_\textup{exp}}\rho_\textup{th}\sqrt{\rho_\textup{exp}}}\right)^2$}
between the reconstructed experimental density matrix $\rho_\textup{exp}$ and the ideal Bell state $\rho_\textup{th}$ is recomputed.

\subsection{SKR Analysis}

In the asymptotic regime, we compute a lower bound on the SKR as \cite{pirandola2020advances}:
\begin{equation}\begin{split}
    \textrm{SKR} \geq [1 - H_2( \mathrm{QBER}_{Z})  f - H_2( \mathrm{QBER}_{X})]  S \, R_{r} \, \alpha \, \eta,
    \label{eq:skr}
\end{split}\end{equation}
where $H_2(x)$ denotes the binary Shannon entropy.
The parameter $f > 1$ is the error reconciliation efficiency, with $H_2(\mathrm{QBER}_{Z}) f$ representing the fraction of bits lost during error correction. The term $H_2(\mathrm{QBER}_{X})$ quantifies the leakage due to privacy amplification, ensuring Eve’s knowledge of the final key is negligible. $S$ is 
the sifting ratio, $R_r$ is the raw bit generation rate, $\alpha$ accounts for transmission losses, and $\eta$ (91\%) is the detectors' efficiency.

Finite-size effects account for the statistical uncertainty in estimating QBERs from finite samples. We bound this uncertainty using the Hoeffding inequality, which provides a confidence interval for the true error rate based on a security parameter $\epsilon_{\text{sec}}$. Taking the worst-case error rates reduces the extractable key rate compared to the asymptotic limit \cite{pirandola2020advances, yin2020entanglement}:

\begin{equation}
\textcolor{black}{\textrm{SKR}_{fin}=\textrm{SKR}-\Delta(n_Z, \epsilon_{\text{sec}})-\lambda_{EV}.}
\label{eq:skr_finite}
\end{equation}
Here, $\Delta(n_Z, \epsilon_{\text{sec}}) = \sqrt{\log(1/\epsilon_{\text{sec}})/n_Z}$ is derived from Hoeffding’s inequality and
quantifies the statistical penalty for finite sample size, with $\epsilon_{\text{sec}} \sim 10^{-12}$ bounding the failure probability.
The term $\lambda_{EV} = \log_2(2/\epsilon_{\text{cor}})/\tau_{n_Z}$ addresses the error verification step, with $\tau_{n_Z}$ being the acquisition time for a block of size $n_Z$, and guarantees that the probability of Alice's and Bob's keys differing is below $\epsilon_{cor} \sim 10^{-12}$.

%% file: text/ackowledgments.tex
\section*{Acknowledgments}
This project has received funding from the European Union’s Horizon Europe Research and Innovation Programme under the QPIC1550 project (101135785) and the European Union’s HORIZON-CL4-2021-DIGITAL-EMERGING-52001 programme under the PROMETHEUS project (101070195). The authors acknowledge financial support from the Carlsberg Foundation and the Novo Nordisk Foundation for the high-efficiency large-scale superconducting detector infrastructure in DTU Electro's SPOC Labs. F.D.R. acknowledges financial support from the Villum Foundation (project OPTIC-AI grant n. VIL29344) D.B. acknowledges support from the European Union ERC StG, QOMUNE (101077917).
M.Z. acknowledges financial support from the European Union’s Digital Europe programme (project EuroQCI, 101091659). The authors would like to thank Christian Koefoed Schou, Silas Lasak Hedeboe and Leif K. Oxenløwe from DTU, and Bera Palsdottir from Lightera Denmark Aps for providing the two-core fiber. This part of the work has been supported by the EDOCS (4354-00020B) IFD funded project.

%% file: text/contributions.tex
\section*{Author contributions}
C.V. and D.B. proposed the experiment and the silicon quantum photonic circuit scheme. C.V. designed the two silicon photonic chips. M.Z. designed the intra-chip communication system. G.G., Da.R. and M.Z. carried out the experiment, collected and analyzed the experimental data. G.G. and Do.R. analyzed the data on the QBER and the SKR. Y.D. and S.Z. fabricated the device. G.G. and Da.R. wrote the manuscript, with input from all authors. C.V., M.Z., B.M., F.D.R., and D.B supervised the whole project.

%% file: bibliography.bib
@article{wang2018multidimensional,
  title={Multidimensional quantum entanglement with large-scale integrated optics},
  author={Wang, Jianwei and Paesani, Stefano and Ding, Yunhong and Santagati, Raffaele and Skrzypczyk, Paul and Salavrakos, Alexia and Tura, Jordi and Augusiak, Remigiusz and Man{\v{c}}inska, Laura and Bacco, Davide and others},
  journal={Science},
  volume={360},
  number={6386},
  pages={285--291},
  year={2018},
  publisher={American Association for the Advancement of Science}
}

@article{adcock2019programmable,
  title={Programmable four-photon graph states on a silicon chip},
  author={Adcock, Jeremy C and Vigliar, Caterina and Santagati, Raffaele and Silverstone, Joshua W and Thompson, Mark G},
  journal={Nature communications},
  volume={10},
  number={1},
  pages={3528},
  year={2019},
  publisher={Nature Publishing Group UK London}
}

@article{vigliar2021error,
  title={Error-protected qubits in a silicon photonic chip},
  author={Vigliar, Caterina and Paesani, Stefano and Ding, Yunhong and Adcock, Jeremy C and Wang, Jianwei and Morley-Short, Sam and Bacco, Davide and Oxenl{\o}we, Leif K and Thompson, Mark G and Rarity, John G and others},
  journal={Nature Physics},
  volume={17},
  number={10},
  pages={1137--1143},
  year={2021},
  publisher={Nature Publishing Group UK London}
}

@article{bao2023very,
  title={Very-large-scale integrated quantum graph photonics},
  author={Bao, Jueming and Fu, Zhaorong and Pramanik, Tanumoy and Mao, Jun and Chi, Yulin and Cao, Yingkang and Zhai, Chonghao and Mao, Yifei and Dai, Tianxiang and Chen, Xiaojiong and others},
  journal={Nature Photonics},
  volume={17},
  number={7},
  pages={573--581},
  year={2023},
  publisher={Nature Publishing Group UK London}
}

@article{ribezzo2022deploying,
  author        = "Ribezzo, Domenico and Zahidy, Mujtaba and Vagniluca, Ilaria and Biagi, Nicola and Francesconi, Saverio and Occhipinti, Tommaso and Bacco, Davide",
  title         = "Deploying an Inter-European Quantum Network",
  journal       = "Advanced Quantum Technologies",
  volume        = "6",
  number        = "2",
  pages         = "2200061",
  year          = "2023",
  doi           = "https://doi.org/10.1002/qute.202200061"
}

@article{scarani2009security,
  title={The security of practical quantum key distribution},
  author={Scarani, Valerio and Bechmann-Pasquinucci, Helle and Cerf, Nicolas J and Du{\v{s}}ek, Miloslav and L{\"u}tkenhaus, Norbert and Peev, Momtchil},
  journal={Reviews of modern physics},
  volume={81},
  number={3},
  pages={1301},
  year={2009},
  publisher={APS}
}

@article{bennett1992quantum,
  title={Quantum cryptography},
  author={Bennett, Charles H and Brassard, Gilles and Ekert, Artur K},
  journal={Scientific American},
  volume={267},
  number={4},
  pages={50--57},
  year={1992},
  publisher={JSTOR}
}

@article{yin2020entanglement,
  title={Entanglement-based secure quantum cryptography over 1,120 kilometres},
  author={Yin, Juan and Li, Yu-Huai and Liao, Sheng-Kai and Yang, Meng and Cao, Yuan and Zhang, Liang and Ren, Ji-Gang and Cai, Wen-Qi and Liu, Wei-Yue and Li, Shuang-Lin and others},
  journal={Nature},
  volume={582},
  number={7813},
  pages={501--505},
  year={2020},
  publisher={Nature Publishing Group}
}

@article{bacco2019field,
  author        = "Bacco, Davide and Vagniluca, Ilaria and Da Lio, Beatrice and Biagi, Nicola and Della Frera, Adriano and Calonico, Davide and Toninelli, Costanza and Cataliotti, Francesco S and Bellini, Marco and Oxenlowe, Leif K and others",
  title         = "Field trial of a three-state quantum key distribution scheme in the Florence metropolitan area",
  journal       = "EPJ Quantum Technology",
  volume        = "6",
  number        = "1",
  pages         = "5",
  year          = "2019",
  doi           = "https://doi.org/10.1140/epjqt/s40507-019-0075-x"
}

@article{llewellyn2020chip,
  author        = "Llewellyn, Daniel and Ding, Yunhong and Faruque, Imad I and Paesani, Stefano and Bacco, Davide and Santagati, Raffaele and Qian, Yan-Jun and Li, Yan and Xiao, Yun-Feng and Huber, Marcus and others",
  title         = "Chip-to-chip quantum teleportation and multi-photon entanglement in silicon",
  journal       = "Nature Physics",
  volume        = "16",
  number        = "2",
  pages         = "148--153",
  year          = "2020",
  doi           = "https://doi.org/10.1038/s41567-019-0727-x"
}

@article{silverstone2015qubit,
  author        = "Silverstone, Joshua W and Santagati, Raffaele and Bonneau, Damien and Strain, Michael J and Sorel, Marc and O’Brien, Jeremy L and Thompson, Mark G",
  title         = "Qubit entanglement between ring-resonator photon-pair sources on a silicon chip",
  journal       = "Nature communications",
  volume        = "6",
  number        = "1",
  pages         = "7948",
  year          = "2015",
  doi           = "https://doi.org/10.1038/ncomms8948"
}

@article{sibson2017chip,
  author        = "Sibson, Philip and Erven, Chris and Godfrey, Mark and Miki, Shigehito and Yamashita, Taro and Fujiwara, Mikio and Sasaki, Masahide and Terai, Hirotaka and Tanner, Michael G and Natarajan, Chandra M and others",
  title         = "Chip-based quantum key distribution",
  journal       = "Nature communications",
  volume        = "8",
  number        = "1",
  pages         = "13984",
  year          = "2017",
  doi           = "https://doi.org/10.1038/ncomms13984"
}

@article{sibson2017integrated,
  author        = "Sibson, Philip and Kennard, Jake E and Stanisic, Stasja and Erven, Chris and O'Brien, Jeremy L and Thompson, Mark G",
  title         = "Integrated silicon photonics for high-speed quantum key distribution",
  journal       = "Optica",
  volume        = "4",
  number        = "2",
  pages         = "172--177",
  year          = "2017",
  doi           = "https://doi.org/10.1364/OPTICA.4.000172"
}

@article{da2021path,
  title={Path-encoded high-dimensional quantum communication over a 2-km multicore fiber},
  author={Da Lio, Beatrice and Cozzolino, Daniele and Biagi, Nicola and Ding, Yunhong and Rottwitt, Karsten and Zavatta, Alessandro and Bacco, Davide and Oxenl{\o}we, Leif K},
  journal={npj Quantum Information},
  volume={7},
  number={1},
  pages={63},
  year={2021},
  publisher={Nature Publishing Group UK London}
}

@article{wang2020integrated,
  title={Integrated photonic quantum technologies},
  author={Wang, Jianwei and Sciarrino, Fabio and Laing, Anthony and Thompson, Mark G},
  journal={Nature Photonics},
  volume={14},
  number={5},
  pages={273--284},
  year={2020},
  publisher={Nature Publishing Group UK London}
}

@article{pirandola2020advances,
  title={Advances in quantum cryptography},
  author={Pirandola, Stefano and Andersen, Ulrik L and Banchi, Leonardo and Berta, Mario and Bunandar, Darius and Colbeck, Roger and Englund, Dirk and Gehring, Tobias and Lupo, Cosmo and Ottaviani, Carlo and others},
  journal={Advances in optics and photonics},
  volume={12},
  number={4},
  pages={1012--1236},
  year={2020},
  publisher={Optica Publishing Group}
}

@article{khodadad2025frequency,
  title={Frequency-bin-encoded entanglement-based quantum key distribution in a reconfigurable frequency-multiplexed network},
  author={Khodadad Kashi, Anahita and Kues, Michael},
  journal={Light: Science \& Applications},
  volume={14},
  number={1},
  pages={49},
  year={2025},
  publisher={Nature Publishing Group UK London}
}

@article{wehner2018quantum,
  title={Quantum internet: A vision for the road ahead},
  author={Wehner, Stephanie and Elkouss, David and Hanson, Ronald},
  journal={Science},
  volume={362},
  number={6412},
  pages={eaam9288},
  year={2018},
  publisher={American Association for the Advancement of Science}
}

@article{neumann2022continuous,
  title={Continuous entanglement distribution over a transnational 248 km fiber link},
  author={Neumann, Sebastian Philipp and Buchner, Alexander and Bulla, Lukas and Bohmann, Martin and Ursin, Rupert},
  journal={Nature Communications},
  volume={13},
  number={1},
  pages={6134},
  year={2022},
  publisher={Nature Publishing Group UK London}
}

@article{xiang2022silicon,
  title={Silicon nitride passive and active photonic integrated circuits: trends and prospects},
  author={Xiang, Chao and Jin, Warren and Bowers, John E},
  journal={Photonics research},
  volume={10},
  number={6},
  pages={A82--A96},
  year={2022},
  publisher={Chinese Laser Press and Optica Publishing Group}
}

@article{ramelow2015silicon,
  title={Silicon-nitride platform for narrowband entangled photon generation},
  author={Ramelow, Sven and Farsi, Alessandro and Clemmen, St{\'e}phane and Orquiza, Daniel and Luke, Kevin and Lipson, Michal and Gaeta, Alexander L},
  journal={arXiv preprint arXiv:1508.04358},
  year={2015}
}

@article{tagliavacche2025frequency,
  title={Frequency-bin entanglement-based quantum key distribution},
  author={Tagliavacche, Noemi and Borghi, Massimo and Guarda, Giulia and Ribezzo, Domenico and Liscidini, Marco and Bacco, Davide and Galli, Matteo and Bajoni, Daniele},
  journal={npj Quantum Information},
  volume={11},
  number={1},
  pages={60},
  year={2025},
  publisher={Nature Publishing Group UK London}
}

@article{azuma2023quantum,
  title={Quantum repeaters: From quantum networks to the quantum internet},
  author={Azuma, Koji and Economou, Sophia E and Elkouss, David and Hilaire, Paul and Jiang, Liang and Lo, Hoi-Kwong and Tzitrin, Ilan},
  journal={Reviews of Modern Physics},
  volume={95},
  number={4},
  pages={045006},
  year={2023},
  publisher={APS}
}

@article{xu2020secure,
  title={Secure quantum key distribution with realistic devices},
  author={Xu, Feihu and Ma, Xiongfeng and Zhang, Qiang and Lo, Hoi-Kwong and Pan, Jian-Wei},
  journal={Reviews of modern physics},
  volume={92},
  number={2},
  pages={025002},
  year={2020},
  publisher={APS}
}

@article{yin2017satellite,
  title={Satellite-to-ground entanglement-based quantum key distribution},
  author={Yin, Juan and Cao, Yuan and Li, Yu-Huai and Ren, Ji-Gang and Liao, Sheng-Kai and Zhang, Liang and Cai, Wen-Qi and Liu, Wei-Yue and Li, Bo and Dai, Hui and others},
  journal={Physical review letters},
  volume={119},
  number={20},
  pages={200501},
  year={2017},
  publisher={APS}
}

@article{fitzke2022scalable,
  title={Scalable network for simultaneous pairwise quantum key distribution via entanglement-based time-bin coding},
  author={Fitzke, Erik and Bialowons, Lucas and Dolejsky, Till and Tippmann, Maximilian and Nikiforov, Oleg and Walther, Thomas and Wissel, Felix and Gunkel, Matthias},
  journal={PRX Quantum},
  volume={3},
  number={2},
  pages={020341},
  year={2022},
  publisher={APS}
}

@article{wen2022realizing,
  title={Realizing an entanglement-based multiuser quantum network with integrated photonics},
  author={Wen, Wenjun and Chen, Zhiyu and Lu, Liangliang and Yan, Wenhan and Xue, Wenyi and Zhang, Peiyu and Lu, Yanqing and Zhu, Shining and Ma, Xiao-song},
  journal={Physical Review Applied},
  volume={18},
  number={2},
  pages={024059},
  year={2022},
  publisher={APS}
}

@article{appas2021flexible,
  title={Flexible entanglement-distribution network with an AlGaAs chip for secure communications},
  author={Appas, F{\'e}licien and Baboux, Florent and Amanti, Maria I and Lema{\'\i}tre, Aristide and Boitier, Fabien and Diamanti, Eleni and Ducci, Sara},
  journal={npj Quantum Information},
  volume={7},
  number={1},
  pages={118},
  year={2021},
  publisher={Nature Publishing Group UK London}
}

@article{steiner2023continuous,
  title={Continuous entanglement distribution from an AlGaAs-on-insulator microcomb for quantum communications},
  author={Steiner, Trevor J and Shen, Maximilian and Castro, Joshua E and Bowers, John E and Moody, Galan},
  journal={Optica Quantum},
  volume={1},
  number={2},
  pages={55--62},
  year={2023},
  publisher={Optica Publishing Group}
}

@article{wang2016chip,
  title={Chip-to-chip quantum photonic interconnect by path-polarization interconversion},
  author={Wang, Jianwei and Bonneau, Damien and Villa, Matteo and Silverstone, Joshua W and Santagati, Raffaele and Miki, Shigehito and Yamashita, Taro and Fujiwara, Mikio and Sasaki, Masahide and Terai, Hirotaka and others},
  journal={Optica},
  volume={3},
  number={4},
  pages={407--413},
  year={2016},
  publisher={Optica Publishing Group}
}

@article{yu2025quantum,
  title={Quantum key distribution implemented with d-level time-bin entangled photons},
  author={Yu, Hao and Sciara, Stefania and Chemnitz, Mario and Montaut, Nicola and Crockett, Benjamin and Fischer, Bennet and Helsten, Robin and Wetzel, Benjamin and Goebel, Thorsten A and Kr{\"a}mer, Ria G and others},
  journal={Nature Communications},
  volume={16},
  number={1},
  pages={171},
  year={2025},
  publisher={Nature Publishing Group UK London}
}

@article{cozzolino2019high,
  title={High-dimensional quantum communication: benefits, progress, and future challenges},
  author={Cozzolino, Daniele and Da Lio, Beatrice and Bacco, Davide and Oxenl{\o}we, Leif Katsuo},
  journal={Advanced Quantum Technologies},
  volume={2},
  number={12},
  pages={1900038},
  year={2019},
  publisher={Wiley Online Library}
}

@article{autebert2016multi,
  title={Multi-user quantum key distribution with entangled photons from an AlGaAs chip},
  author={Autebert, Claire and Trapateau, Julien and Orieux, Adeline and Lema{\^\i}tre, Aristide and Gomez-Carbonell, Carmen and Diamanti, Eleni and Zaquine, Isabelle and Ducci, Sara},
  journal={Quantum Science and Technology},
  volume={1},
  number={1},
  pages={01LT02},
  year={2016},
  publisher={IOP Publishing}
}

@article{liu2023photonic,
  title={Photonic-reconfigurable entanglement distribution network based on silicon quantum photonics},
  author={Liu, Dongning and Liu, Jingyuan and Ren, Xiaosong and Feng, Xue and Liu, Fang and Cui, Kaiyu and Huang, Yidong and Zhang, Wei},
  journal={Photonics Research},
  volume={11},
  number={7},
  pages={1314--1325},
  year={2023},
  publisher={Optica Publishing Group}
}

@article{semenenko2020chip,
  title={Chip-based measurement-device-independent quantum key distribution},
  author={Semenenko, Henry and Sibson, Philip and Hart, Andy and Thompson, Mark G and Rarity, John G and Erven, Chris},
  journal={Optica},
  volume={7},
  number={3},
  pages={238--242},
  year={2020},
  publisher={Optical Society of America}
}

@article{politi2008silica,
  title={Silica-on-silicon waveguide quantum circuits},
  author={Politi, Alberto and Cryan, Martin J and Rarity, John G and Yu, Siyuan and O'brien, Jeremy L},
  journal={Science},
  volume={320},
  number={5876},
  pages={646--649},
  year={2008},
  publisher={American Association for the Advancement of Science}
}

@article{shadbolt2012generating,
  title={Generating, manipulating and measuring entanglement and mixture with a reconfigurable photonic circuit},
  author={Shadbolt, Peter J and Verde, Maria R and Peruzzo, Alberto and Politi, Alberto and Laing, Anthony and Lobino, Mirko and Matthews, Jonathan CF and Thompson, Mark G and O'Brien, Jeremy L},
  journal={Nature Photonics},
  volume={6},
  number={1},
  pages={45--49},
  year={2012},
  publisher={Nature Publishing Group UK London}
}

@article{silverstone2014chip,
  title={On-chip quantum interference between silicon photon-pair sources},
  author={Silverstone, Joshua W and Bonneau, Damien and Ohira, Kazuya and Suzuki, Nob and Yoshida, Haruhiko and Iizuka, Norio and Ezaki, Mizunori and Natarajan, Chandra M and Tanner, Michael G and Hadfield, Robert H and others},
  journal={Nature Photonics},
  volume={8},
  number={2},
  pages={104--108},
  year={2014},
  publisher={Nature Publishing Group UK London}
}

@article{tanzilli2012genesis,
  title={On the genesis and evolution of integrated quantum optics},
  author={Tanzilli, S{\'e}bastien and Martin, Anthony and Kaiser, Florian and De Micheli, Marc P and Alibart, Olivier and Ostrowsky, Daniel B},
  journal={Laser \& Photonics Reviews},
  volume={6},
  number={1},
  pages={115--143},
  year={2012},
  publisher={Wiley Online Library}
}

@article{zhang2019generation,
  title={Generation of multiphoton quantum states on silicon},
  author={Zhang, Ming and Feng, Lan-Tian and Zhou, Zhi-Yuan and Chen, Yang and Wu, Hao and Li, Ming and Gao, Shi-Ming and Guo, Guo-Ping and Guo, Guang-Can and Dai, Dao-Xin and others},
  journal={Light: Science \& Applications},
  volume={8},
  number={1},
  pages={41},
  year={2019},
  publisher={Nature Publishing Group UK London}
}

@article{paesani2019generation,
  title={Generation and sampling of quantum states of light in a silicon chip},
  author={Paesani, Stefano and Ding, Yunhong and Santagati, Raffaele and Chakhmakhchyan, Levon and Vigliar, Caterina and Rottwitt, Karsten and Oxenl{\o}we, Leif K and Wang, Jianwei and Thompson, Mark G and Laing, Anthony},
  journal={Nature Physics},
  volume={15},
  number={9},
  pages={925--929},
  year={2019},
  publisher={Nature Publishing Group UK London}
}

@article{metcalf2014quantum,
  title={Quantum teleportation on a photonic chip},
  author={Metcalf, Benjamin J and Spring, Justin B and Humphreys, Peter C and Thomas-Peter, Nicholas and Barbieri, Marco and Kolthammer, W Steven and Jin, Xian-Min and Langford, Nathan K and Kundys, Dmytro and Gates, James C and others},
  journal={Nature photonics},
  volume={8},
  number={10},
  pages={770--774},
  year={2014},
  publisher={Nature Publishing Group UK London}
}

@article{ren2017ground,
  title={Ground-to-satellite quantum teleportation},
  author={Ren, Ji-Gang and Xu, Ping and Yong, Hai-Lin and Zhang, Liang and Liao, Sheng-Kai and Yin, Juan and Liu, Wei-Yue and Cai, Wen-Qi and Yang, Meng and Li, Li and others},
  journal={Nature},
  volume={549},
  number={7670},
  pages={70--73},
  year={2017},
  publisher={Nature Publishing Group UK London}
}

@inproceedings{guarda2023bb84,
  title={{BB84} decoy-state {QKD} protocol over long-distance optical fiber},
  author={Guarda, Giulia and Ribezzo, Domenico and Salvoni, Daniela and Bruscino, Ciro and Ercolano, Pasquale and Ejrnaes, Mikkel and Parlato, Loredana and Zhang, C and Li, H and You, L and others},
  booktitle={2023 23rd International Conference on Transparent Optical Networks (ICTON)},
  pages={1--4},
  year={2023},
  organization={IEEE}
}

@article{ding2017high,
  title={High-dimensional quantum key distribution based on multicore fiber using silicon photonic integrated circuits},
  author={Ding, Yunhong and Bacco, Davide and Dalgaard, Kjeld and Cai, Xinlun and Zhou, Xiaoqi and Rottwitt, Karsten and Oxenl{\o}we, Leif Katsuo},
  journal={npj Quantum Information},
  volume={3},
  number={1},
  pages={25},
  year={2017},
  publisher={Nature Publishing Group UK London}
}

@article{da2019stable,
  title={Stable transmission of high-dimensional quantum states over a 2-km multicore fiber},
  author={Da Lio, Beatrice and Bacco, Davide and Cozzolino, Daniele and Biagi, Nicola and Arge, Tummas Napoleon and Larsen, Emil and Rottwitt, Karsten and Ding, Yunhong and Zavatta, Alessandro and Oxenl{\o}we, Leif Katsuo},
  journal={IEEE Journal of Selected Topics in Quantum Electronics},
  volume={26},
  number={4},
  pages={1--8},
  year={2019},
  publisher={IEEE}
}

@article{wang2024ultrabroadband,
  title={Ultrabroadband thin-film lithium tantalate modulator for high-speed communications},
  author={Wang, Chengli and Fang, Dengyang and Zhang, Junyin and Kotz, Alexander and Lihachev, Grigory and Churaev, Mikhail and Li, Zihan and Schwarzenberger, Adrian and Ou, Xin and Koos, Christian and others},
  journal={Optica},
  volume={11},
  number={12},
  pages={1614--1620},
  year={2024},
  publisher={Optica Publishing Group}
}

@article{xu2020high,
  title={High-performance coherent optical modulators based on thin-film lithium niobate platform},
  author={Xu, Mengyue and He, Mingbo and Zhang, Hongguang and Jian, Jian and Pan, Ying and Liu, Xiaoyue and Chen, Lifeng and Meng, Xiangyu and Chen, Hui and Li, Zhaohui and others},
  journal={Nature communications},
  volume={11},
  number={1},
  pages={3911},
  year={2020},
  publisher={Nature Publishing Group UK London}
}

@article{krenn2017entanglement,
  title={Entanglement by path identity},
  author={Krenn, Mario and Hochrainer, Armin and Lahiri, Mayukh and Zeilinger, Anton},
  journal={Physical review letters},
  volume={118},
  number={8},
  pages={080401},
  year={2017},
  publisher={APS}
}

@article{ma2016silicon,
  title={Silicon photonic transmitter for polarization-encoded quantum key distribution},
  author={Ma, Chaoxuan and Sacher, Wesley D and Tang, Zhiyuan and Mikkelsen, Jared C and Yang, Yisu and Xu, Feihu and Thiessen, Torrey and Lo, Hoi-Kwong and Poon, Joyce KS},
  journal={Optica},
  volume={3},
  number={11},
  pages={1274--1278},
  year={2016},
  publisher={OSA}
}

@misc{thomas2025high,
  author = {Thomas,  Molly A. and Llewellyn,  Daniel and Yard,  Patrick W. and Slater,  Benjamin A. and Vigliar,  Caterina and Paesani,  Stefano and Borghi,  Massimo and Sahin,  D\"{o}nd\"{u} and Rarity,  John G. and Oxenløwe,  Leif K. and Thompson,  Mark G. and Rottwitt,  Karsten and Ding,  Yunhong and Wang,  Jianwei and Bacco,  Davide and Barreto,  Jorge},
  title = {High-dimensional Path-Encoded Entanglement Distribution Between Photonic Chips Enabled by Multimode Phase Stabilisation},
  publisher = {arXiv},
  year = {2025},
  copyright = {Creative Commons Attribution 4.0 International}
}

@article{ding2014fully,
  title={Fully etched apodized grating coupler on the {SOI} platform with- 0.58 dB coupling efficiency},
  author={Ding, Yunhong and Peucheret, Christophe and Ou, Haiyan and Yvind, Kresten},
  journal={Optics letters},
  volume={39},
  number={18},
  pages={5348--5350},
  year={2014},
  publisher={Optical Society of America}
}

@article{paesani2020near,
  title={Near-ideal spontaneous photon sources in silicon quantum photonics},
  author={Paesani, Stefano and Borghi, Massimo and Signorini, Stefano and Ma{\"\i}nos, Alexandre and Pavesi, Lorenzo and Laing, Anthony},
  journal={Nature communications},
  volume={11},
  number={1},
  pages={2505},
  year={2020},
  publisher={Nature Publishing Group UK London}
}

@article{James2001,
  title = {Measurement of qubits},
  volume = {64},
  ISSN = {1094-1622},
  DOI = {10.1103/physreva.64.052312},
  number = {5},
  journal = {Physical Review A},
  publisher = {American Physical Society (APS)},
  author = {James,  Daniel F. V. and Kwiat,  Paul G. and Munro,  William J. and White,  Andrew G.},
  year = {2001},
  month = oct 
}

@article{Zahidy2024,
  title = {Quantum key distribution using deterministic single-photon sources over a field-installed fibre link},
  volume = {10},
  ISSN = {2056-6387},
  DOI = {10.1038/s41534-023-00800-x},
  number = {1},
  journal = {npj Quantum Information},
  publisher = {Springer Science and Business Media LLC},
  author = {Zahidy,  Mujtaba and Mikkelsen,  Mikkel T. and M\"{u}ller,  Ronny and Da Lio,  Beatrice and Krehbiel,  Martin and Wang,  Ying and Bart,  Nikolai and Wieck,  Andreas D. and Ludwig,  Arne and Galili,  Michael and Forchhammer,  Søren and Lodahl,  Peter and Oxenløwe,  Leif K. and Bacco,  Davide and Midolo,  Leonardo},
  year = {2024},
  month = jan 
}

@article{Xiong2016,
  title = {Active temporal multiplexing of indistinguishable heralded single photons},
  volume = {7},
  ISSN = {2041-1723},
  DOI = {10.1038/ncomms10853},
  number = {1},
  journal = {Nature Communications},
  publisher = {Springer Science and Business Media LLC},
  author = {Xiong,  C. and Zhang,  X. and Liu,  Z. and Collins,  M. J. and Mahendra,  A. and Helt,  L. G. and Steel,  M. J. and Choi,  D. -Y. and Chae,  C. J. and Leong,  P. H. W. and Eggleton,  B. J.},
  year = {2016},
  month = mar 
}

@article{Davis2022,
  title = {Improved Heralded Single-Photon Source with a Photon-Number-Resolving Superconducting Nanowire Detector},
  volume = {18},
  ISSN = {2331-7019},
  DOI = {10.1103/physrevapplied.18.064007},
  number = {6},
  journal = {Physical Review Applied},
  publisher = {American Physical Society (APS)},
  author = {Davis,  Samantha I. and Mueller,  Andrew and Valivarthi,  Raju and Lauk,  Nikolai and Narvaez,  Lautaro and Korzh,  Boris and Beyer,  Andrew D. and Cerri,  Olmo and Colangelo,  Marco and Berggren,  Karl K. and Shaw,  Matthew D. and Xie,  Si and Sinclair,  Neil and Spiropulu,  Maria},
  year = {2022},
  month = dec 
}

@article{bacco2021mcf,
  title = {Characterization and stability measurement of deployed multicore fibers for quantum applications},
  volume = {9},
  ISSN = {2327-9125},
  number = {10},
  journal = {Photonics Research},
  publisher = {Optica Publishing Group},
  author = {Bacco,  Davide and Biagi,  Nicola and Vagniluca,  Ilaria and Hayashi,  Tetsuya and Mecozzi,  Antonio and Antonelli,  Cristian and Oxenløwe,  Leif K. and Zavatta,  Alessandro},
  year = {2021},
  month = sep,
  pages = {1992}
}

@article{wengerowsky2019submarine,
  title = {Entanglement distribution over a 96-km-long submarine optical fiber},
  volume = {116},
  ISSN = {1091-6490},
  number = {14},
  journal = {Proceedings of the National Academy of Sciences},
  publisher = {Proceedings of the National Academy of Sciences},
  author = {Wengerowsky,  S\"{o}ren and Joshi,  Siddarth Koduru and Steinlechner,  Fabian and Zichi,  Julien R. and Dobrovolskiy,  Sergiy M. and van der Molen,  René and Los,  Johannes W. N. and Zwiller,  Val and Versteegh,  Marijn A. M. and Mura,  Alberto and Calonico,  Davide and Inguscio,  Massimo and H\"{u}bel,  Hannes and Bo,  Liu and Scheidl,  Thomas and Zeilinger,  Anton and Xuereb,  André and Ursin,  Rupert},
  year = {2019},
  month = mar,
  pages = {6684–6688}
}

@article{chi2022programmable,
  title = {A programmable qudit-based quantum processor},
  volume = {13},
  ISSN = {2041-1723},
  number = {1},
  journal = {Nature Communications},
  publisher = {Springer Science and Business Media LLC},
  author = {Chi,  Yulin and Huang,  Jieshan and Zhang,  Zhanchuan and Mao,  Jun and Zhou,  Zinan and Chen,  Xiaojiong and Zhai,  Chonghao and Bao,  Jueming and Dai,  Tianxiang and Yuan,  Huihong and Zhang,  Ming and Dai,  Daoxin and Tang,  Bo and Yang,  Yan and Li,  Zhihua and Ding,  Yunhong and Oxenløwe,  Leif K. and Thompson,  Mark G. and O’Brien,  Jeremy L. and Li,  Yan and Gong,  Qihuang and Wang,  Jianwei},
  year = {2022},
  month = mar 
}

@article{jorgensen2022petabit,
  title = {Petabit-per-second data transmission using a chip-scale microcomb ring resonator source},
  volume = {16},
  ISSN = {1749-4893},
  number = {11},
  journal = {Nature Photonics},
  publisher = {Springer Science and Business Media LLC},
  author = {Jørgensen,  A. A. and Kong,  D. and Henriksen,  M. R. and Klejs,  F. and Ye,  Z. and Helgason,  Ò. B. and Hansen,  H. E. and Hu,  H. and Yankov,  M. and Forchhammer,  S. and Andrekson,  P. and Larsson,  A. and Karlsson,  M. and Schr\"{o}der,  J. and Sasaki,  Y. and Aikawa,  K. and Thomsen,  J. W. and Morioka,  T. and Galili,  M. and Torres-Company,  V. and Oxenløwe,  L. K.},
  year = {2022},
  month = oct,
  pages = {798–802}
}

@article{bogaerts2020programmable,
  title = {Programmable photonic circuits},
  volume = {586},
  ISSN = {1476-4687},

  number = {7828},
  journal = {Nature},
  publisher = {Springer Science and Business Media LLC},
  author = {Bogaerts,  Wim and Pérez,  Daniel and Capmany,  José and Miller,  David A. B. and Poon,  Joyce and Englund,  Dirk and Morichetti,  Francesco and Melloni,  Andrea},
  year = {2020},
  month = oct,
  pages = {207–216}
}

@article{sun2015microprocessor,
  title = {Single-chip microprocessor that communicates directly using light},
  volume = {528},
  ISSN = {1476-4687},
  number = {7583},
  journal = {Nature},
  publisher = {Springer Science and Business Media LLC},
  author = {Sun,  Chen and Wade,  Mark T. and Lee,  Yunsup and Orcutt,  Jason S. and Alloatti,  Luca and Georgas,  Michael S. and Waterman,  Andrew S. and Shainline,  Jeffrey M. and Avizienis,  Rimas R. and Lin,  Sen and Moss,  Benjamin R. and Kumar,  Rajesh and Pavanello,  Fabio and Atabaki,  Amir H. and Cook,  Henry M. and Ou,  Albert J. and Leu,  Jonathan C. and Chen,  Yu-Hsin and Asanović,  Krste and Ram,  Rajeev J. and Popović,  Miloš A. and Stojanović,  Vladimir M.},
  year = {2015},
  month = dec,
  pages = {534–538}
}

@article{margalit2021perspective,
  title = {Perspective on the future of silicon photonics and electronics},
  volume = {118},
  ISSN = {1077-3118},
  DOI = {10.1063/5.0050117},
  number = {22},
  journal = {Applied Physics Letters},
  publisher = {AIP Publishing},
  author = {Margalit,  Near and Xiang,  Chao and Bowers,  Steven M. and Bjorlin,  Alexis and Blum,  Robert and Bowers,  John E.},
  year = {2021},
  month = may 
}

@article{atabaki2018integrating,
  title = {Integrating photonics with silicon nanoelectronics for the next generation of systems on a chip},
  volume = {556},
  ISSN = {1476-4687},
  DOI = {10.1038/s41586-018-0028-z},
  number = {7701},
  journal = {Nature},
  publisher = {Springer Science and Business Media LLC},
  author = {Atabaki,  Amir H. and Moazeni,  Sajjad and Pavanello,  Fabio and Gevorgyan,  Hayk and Notaros,  Jelena and Alloatti,  Luca and Wade,  Mark T. and Sun,  Chen and Kruger,  Seth A. and Meng,  Huaiyu and Al Qubaisi,  Kenaish and Wang,  Imbert and Zhang,  Bohan and Khilo,  Anatol and Baiocco,  Christopher V. and Popović,  Miloš A. and Stojanović,  Vladimir M. and Ram,  Rajeev J.},
  year = {2018},
  month = apr,
  pages = {349–354}
}

@article{Thomson2011,
  title = {High contrast {40Gbit/s} optical modulation in silicon},
  volume = {19},
  ISSN = {1094-4087},
  DOI = {10.1364/oe.19.011507},
  number = {12},
  journal = {Optics Express},
  publisher = {Optica Publishing Group},
  author = {Thomson,  D. J. and Gardes,  F. Y. and Hu,  Y. and Mashanovich,  G. and Fournier,  M. and Grosse,  P. and Fedeli,  J-M. and Reed,  G. T.},
  year = {2011},
  month = May,
  pages = {11507}
}

@article{Grillanda2014,
  title = {Non-invasive monitoring and control in silicon photonics using {CMOS} integrated electronics},
  volume = {1},
  ISSN = {2334-2536},
  DOI = {10.1364/optica.1.000129},
  number = {3},
  journal = {Optica},
  publisher = {Optica Publishing Group},
  author = {Grillanda,  Stefano and Carminati,  Marco and Morichetti,  Francesco and Ciccarella,  Pietro and Annoni,  Andrea and Ferrari,  Giorgio and Strain,  Michael and Sorel,  Marc and Sampietro,  Marco and Melloni,  Andrea},
  year = {2014},
  month = Aug,
  pages = {129}
}

@article{Mateo2024,
  title = {Multicore-fiber submarine systems [Invited]},
  volume = {16},
  ISSN = {1943-0639},
  DOI = {10.1364/jocn.532163},
  number = {11},
  journal = {Journal of Optical Communications and Networking},
  publisher = {Optica Publishing Group},
  author = {Mateo,  Eduardo F.},
  year = {2024},
  month = Oct,
  pages = {H1}
}
